\documentclass[a4paper,11pt]{article}
\usepackage{jinstpub} 
\usepackage{lineno}
\usepackage{float}  

\title{\boldmath ALICE ITS2: overview and performance}

\author{A. Isakov, {on behalf of the ALICE  collaboration}}
\affiliation{Nikhef, Netherlands}

\emailAdd{aisakov@nikhef.nl}

\abstract{The new Inner Tracking System (ITS2) is instrumental for tracking and vertex reconstruction
in the ALICE experiment. The new tracker consists of seven cylindrical layers equipped with silicon
Monolithic Active Pixel Sensors (MAPS) with a pixel size of $27 \times 29\,\mathrm{\mu m}$. The sensors are thinned
down to a thickness of 50 $\mathrm{\mu m}$ and 100 $\mathrm{\mu m}$ for the three innermost layers and for the four outer layers,
respectively. The material budget of the innermost layers is as low as 0.36\% $X_{0}$ per layer compared to
1.14\% $X_{0}$/ layer of the previous ITS1. In combination with a radius of 23 mm for the innermost layer
and a position resolution of about 5 $\mathrm{\mu m}$, the low material budget greatly enhances the reconstruction
capabilities of heavy-flavour and low-$p_{\rm T}$ particles compared to Run 2. 

ITS2 has been in operation since the beginning of Run 3 and has already recorded more than 80 pb$^{-1}$ proton-proton (pp) events at $\sqrt{s}$ = 13.6 TeV and more than 2 nb$^{-1}$ Pb--Pb collisions at  $ \sqrt{s_{ \rm{NN} }} $ = 5.36 TeV, operating stably during these operations at interaction rate up to 4 MHz in pp and about 50 kHz in Pb--Pb collisions. 

This contribution will review the detector performance during LHC Run 3 and give an overview on the calibration methods and the data-taking experience.}


\begin{document}
\maketitle
\flushbottom

\section{Inner Tracking System for Run 3 upgrade}
\label{sec:intro}

A Large Ion Collider Experiment (ALICE) is the heavy-ion focused experiment at the Large Hadron Collider (LHC), designed to investigate the properties of quark-gluon plasma in heavy ion collisions. The current generation of the ALICE Inner Tracking System (ITS2) was installed during the LHC Long Shutdown 2 as part of a detector upgrade to enable recording of all Pb--Pb events at an interaction rate of up to 50 kHz in Run 3 \cite{LS2}. The ITS2 detector targets at a five-fold improvement in position resolution along the $z$-axis and a three-fold enhancement in the $r\varphi$-direction, in addition to up to 8 times increased tracking efficiency for particles with momentum below 1 GeV/$c$ \cite{its2}. 

\begin{figure}[htbp]
\centering
\includegraphics[width=.46\textwidth]{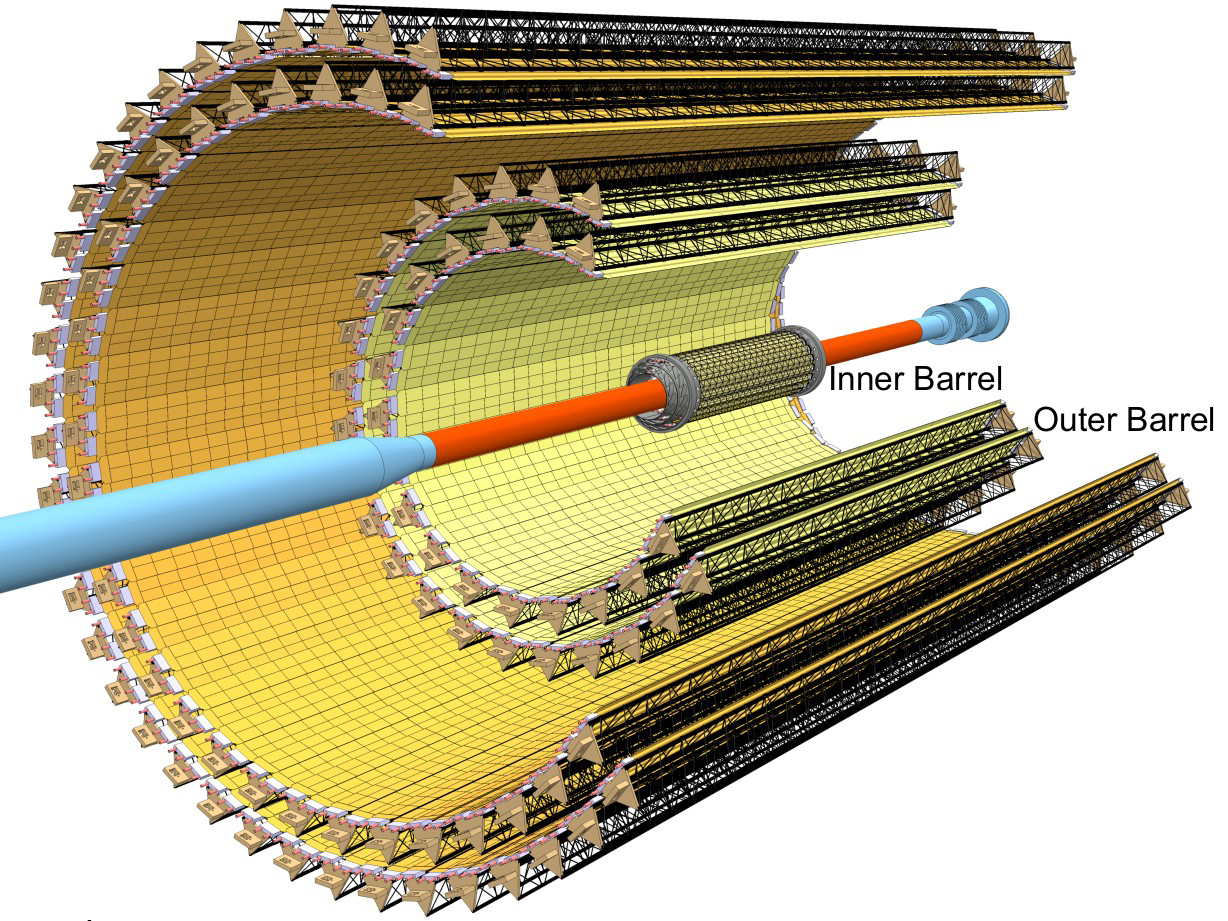}
\qquad
\includegraphics[width=.48\textwidth]{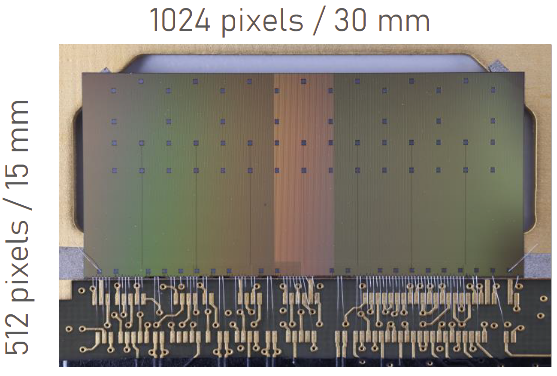}
\caption{Left: scheme of the Upgraded ITS system. Right: ALPIDE sensor.}
\label{fig:its}
\end{figure}

The schematic view of ITS2 is depicted in Figure \ref{fig:its} left and it comprises seven cylindrical layers of Monolithic Active Pixel Sensors (MAPS), encompassing a total active silicon area of 10 m$^2$ and $12.5 \cdot 10^9$ pixels. The first three innermost layers constitute the Inner Barrel (IB), which features a lower material budget per layer ($0.36\%$ X$_0$). Each IB stave consists of single Hybrid Integrated Circuit (HIC) integrating nine ALPIDE chips, with data readout performed through high-speed links operating at 1.2 Gbit/s per chip. Layers 4 to 7 constitute the Outer Barrel (OB), where each stave employs multiple HICs constructed from two rows of seven chips, each 100 $\mathrm{\mu m}$ thick. Each row features a 400 Mbit/s readout capacity \cite{its2}.

The fundamental component of the ITS2 is the ALPIDE sensor (see Figure \ref{fig:its} right), performing in-pixel amplification, signal shaping, and discrimination. The key focus of the chip design was to ensure low power consumption (<47 mW/cm$^2$), alongside with high radiation tolerance for MAPS (>270 kRad of TID, $1.7 \cdot 10^{12}$ 1 MeV $\rm{n_{eq}}$ cm$^{-2}$ of NIEL). These objectives were achieved using 180 nm CMOS pixel technology with a deep p-well imaging process provided by TowerJazz. The pixel design incorporated a small 2 $\mathrm{\mu m}$ diameter n-well diode with low capacitance ($\sim$fF) and a 25 $\mathrm{\mu m}$-thick p-type epitaxial layer \cite{alpide}.

\section{ITS performance in Run 3}
\label{sec:run3}

Since  the official start of Run 3 on July 5th 2022, ITS2 has accumulated over 2,000 hours of data-taking, recording more than 82 pb$^{-1}$ of pp collisions and 2.16 nb$^{-1}$ of Pb--Pb collision data during this period. It has already exceeded the total Pb--Pb statistics acquired during Runs 1 and 2, which amounted to 1.5 nb$^{-1}$. The ITS2 has continuously operated 24,000 chips with an average per-layer run-time impacted by ITS chip issues remaining below 1\%. Moreover, the stability of chip operations is monitored with high precision, down to individual ALPIDE chips, which serves to reproduce the detector performance in Monte-Carlo simulations.

This level of performance has been achieved through the deployment of an advanced auto-recovery system. This system detects and addresses issues encountered during data-taking, such as corrupted data or radiation-induced errors in readout unit firmware and data path, and executes reconfiguration of hardware components, thereby ensuring continuous optimal operation.

\subsection{Tracking performance}

\begin{figure}[htbp]
\centering
\includegraphics[width=.46\textwidth]{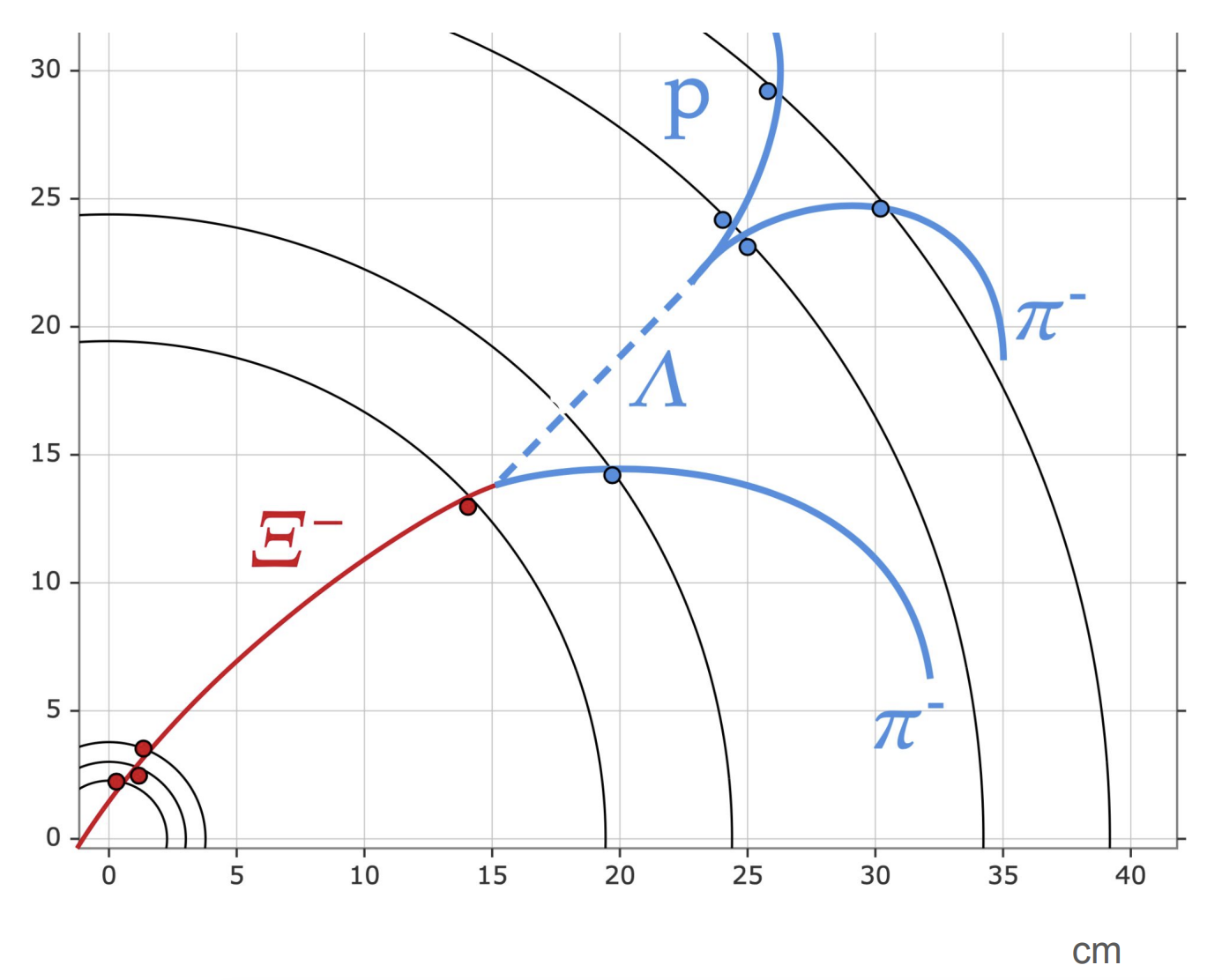}
\qquad
\includegraphics[width=.48\textwidth]{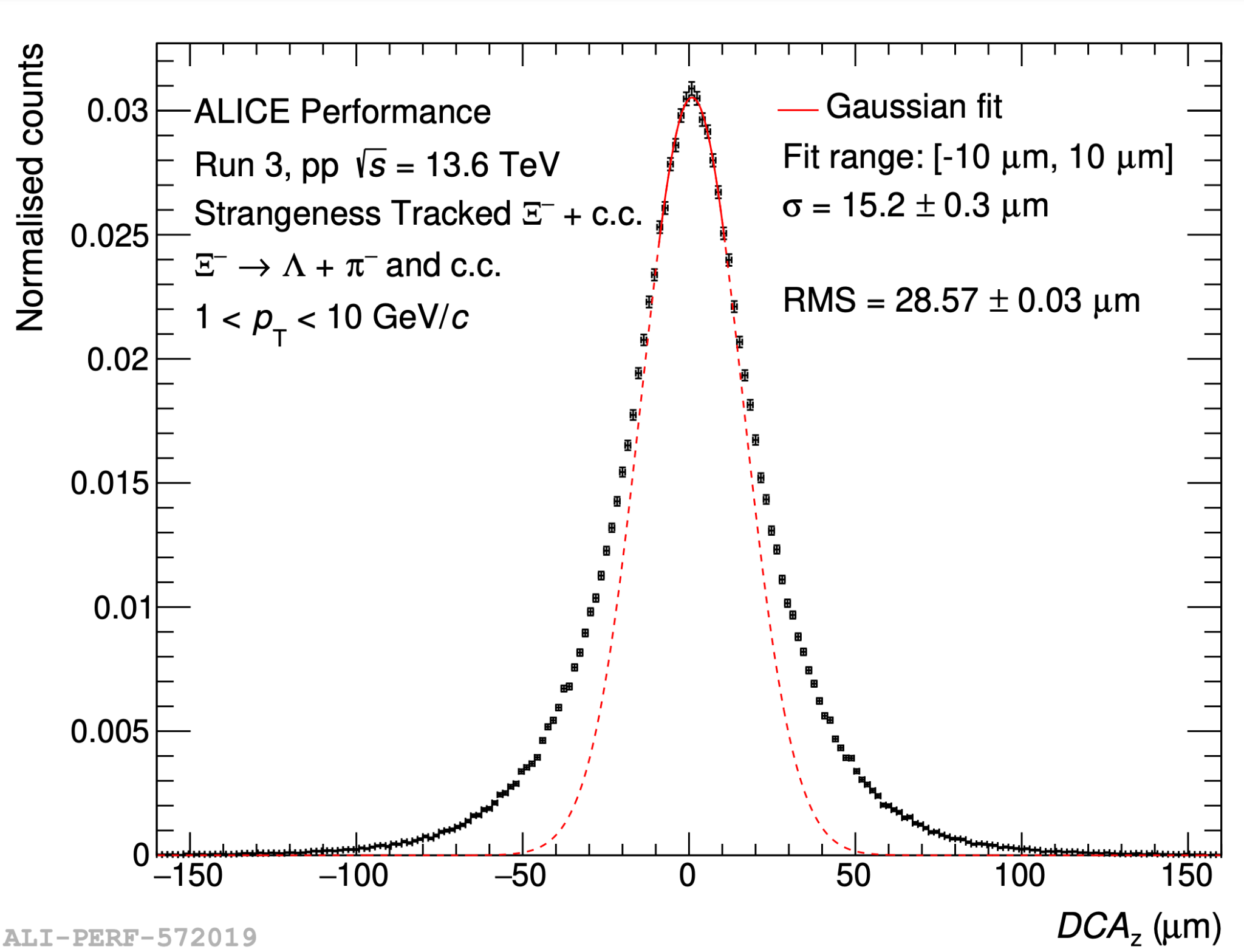}
\caption{Left: scheme of the $\Xi^- \rightarrow \Lambda + \pi^-$ weak decay overlaid on the ITS2 geometry, represented by circular arcs. Right: distance of closest approach (DCA) along the z-axis of the strange baryon, together with a Gaussian fit at $| DCA_z |< 10$  $\mathrm{\mu m}$. The difference between the sigma of the fit and the RMS of the data can be explained by the fact that the RMS estimation includes all non-prompt components, while the Gaussian fit was performed only within a restricted range.}
\label{fig:tracking}
\end{figure}

One of the significant improvements in the ITS2 design is the closer positioning of the innermost layer to the interaction point at 2.2 cm instead of previous 4 cm \cite{its2}. This modification not only enhances the reconstruction efficiency and pointing resolution at low transverse momentum but also enables ITS-standalone tracking of particles before they undergo weak decay. As illustrated by the sketch in the left panel of Figure \ref{fig:tracking}, ITS2 geometry opens the possibility for new measurements of non-prompt cascades with the ITS, such as the weak decay channel of $\Xi^-$ with the total RMS of DCA$_z$ of $28.57 \pm 0.03$ $\mathrm{\mu m}$ (see the right panel of Figure \ref{fig:tracking}).

\subsection{Noise level}

\begin{figure}[htbp]
\centering
\includegraphics[width=.99\textwidth]{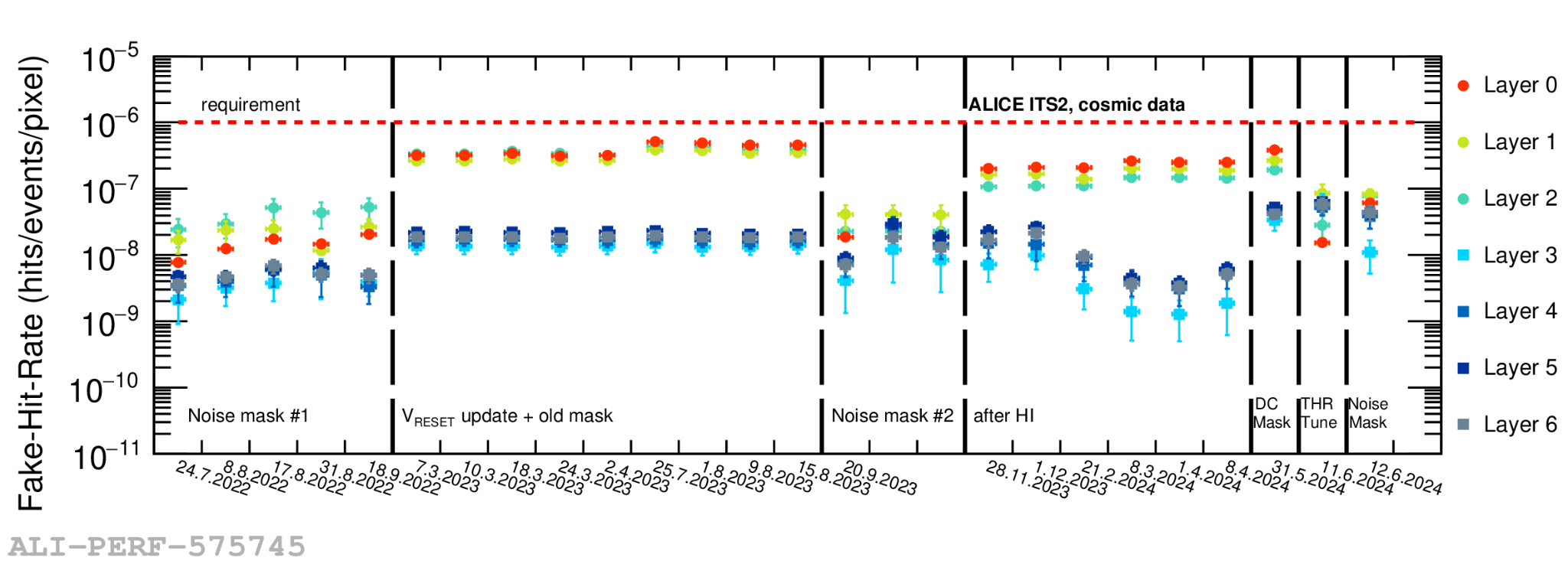}
\caption{Averaged per layer FHR trend as a function of date. Each data point on x-axis corresponds to cosmic run longer than 30 minutes. The ITS2 Upgrade project FHR requirement of $<10^{-6}$ hits per event per pixel for both IB and OB is shown by the red dashed line. Taken from \cite{A_triolo_calib}}
\label{fig:fhr}
\end{figure}

The noise level of the ALPIDE sensors is defined by the fake-hit rate (FHR) measured during data-taking in absence of beam-induced collisions. As shown in the long-term trend on Figure \ref{fig:fhr}, the monitored value consistently remains at least by a factor of ten below the project requirement of 10$^{-6}$ hits/event/pixel. Observed fluctuations are caused by the emergence of new noisy pixels and changes in threshold values; however, these are effectively managed through noisy pixel masking and threshold re-tuning. Based on the latest noisy pixel scan, only 0.15\% of the pixels across the entire detector were masked. It is also noteworthy that stricter noise criteria are applied to OB chips, while detection efficiency is prioritized for IB chips ($>10^{-2}$ and $10^{-6}$ of hits/event/pixel for IB and OB, respectively).

\subsection{Charge threshold instability}

\begin{figure}[htbp]
\centering
\includegraphics[width=.95\textwidth]{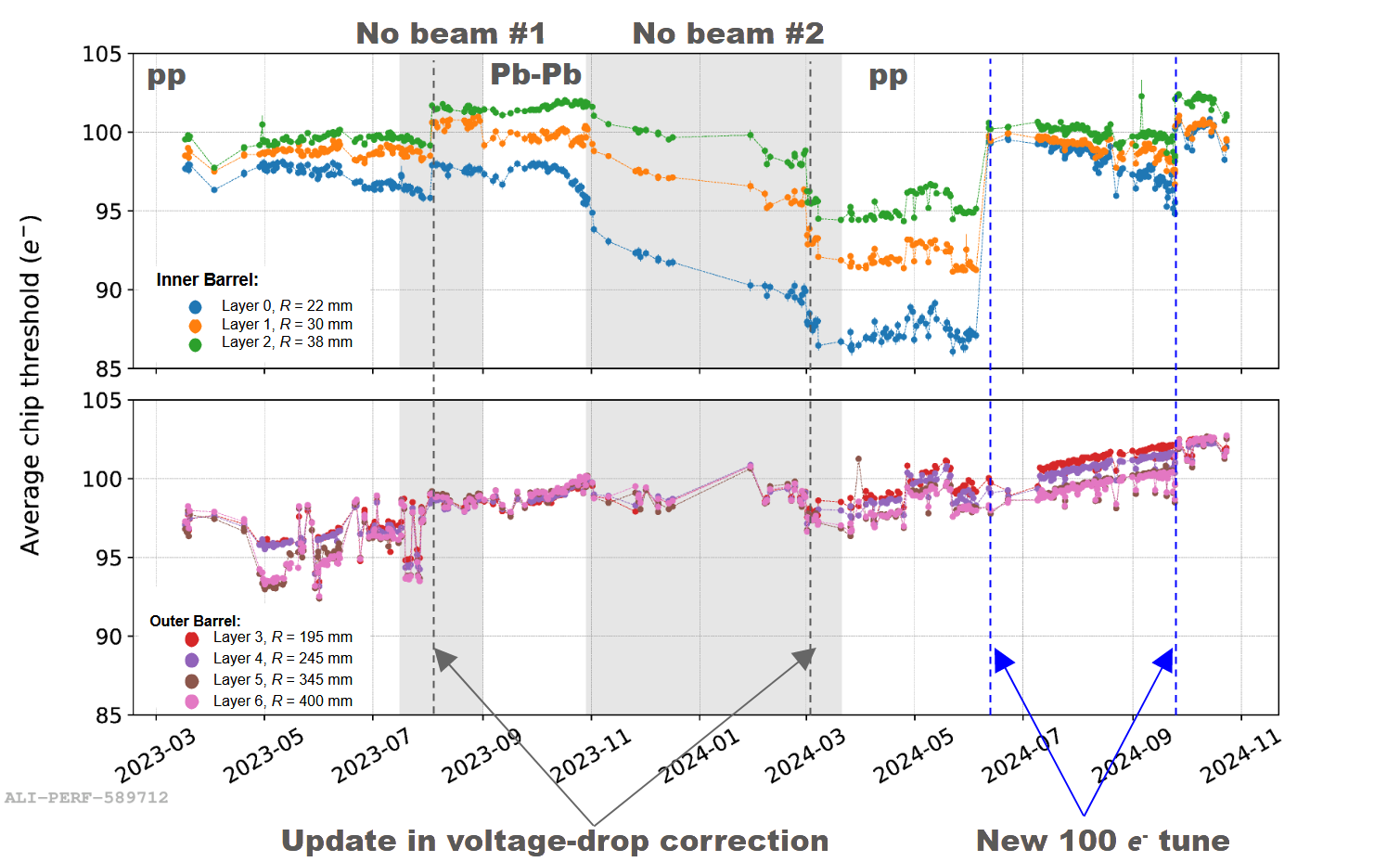}
\caption{Evolution of the averaged per-layer in-pixel threshold from March 2023 to July 2024. Periods without beam-induced collisions are indicated by gray areas, dark gray arrows mark changes in voltage operations, and blue arrows indicate threshold re-tuning. }
\label{fig:thr}
\end{figure}

The charge threshold of the detector is monitored in between accelerator fills. The corresponding trend is shown in the left panel of Figure \ref{fig:thr}. The target charge threshold of 100 $e^-$ is a compromise between detection efficiency (>99\%) and fake-hit rate ($<10^{-6}$ hits/event/pixel) \cite{radiation_efficiency}. The threshold value is sensitive to variations in temperature and analogue supply voltage, which explains the minor fluctuations observed between consecutive scans and the more pronounced jumps during no-beam periods (gray areas), when the voltage-drop correction algorithm was updated. Initially tuned to 100 $e^-$, the threshold level exhibited a slight decrease in the inner layers, starting mid-way through the Pb--Pb run  in 2023. The magnitude of this decrease correlates with the accumulated radiation dose and is inversely proportional to the radial distance squared from the collision point. By February 2024, this behavior had stabilized at a threshold value of 85 $e^-$, after which the detector was re-tuned to the nominal charge threshold values. During this period, no decrease in detector efficiency was observed, and the FHR remained within acceptable limits \cite{radiation_efficiency}. Overall, the observed threshold behavior is consistent with radiation hardness studies of 180 nm CMOS MAPS test structures carried out during the ITS2 R\&D phase \cite{hillemanns2013}.

\subsection{Particle identification with MAPS}

\begin{figure}[H]
\centering
\includegraphics[width=.75\textwidth]{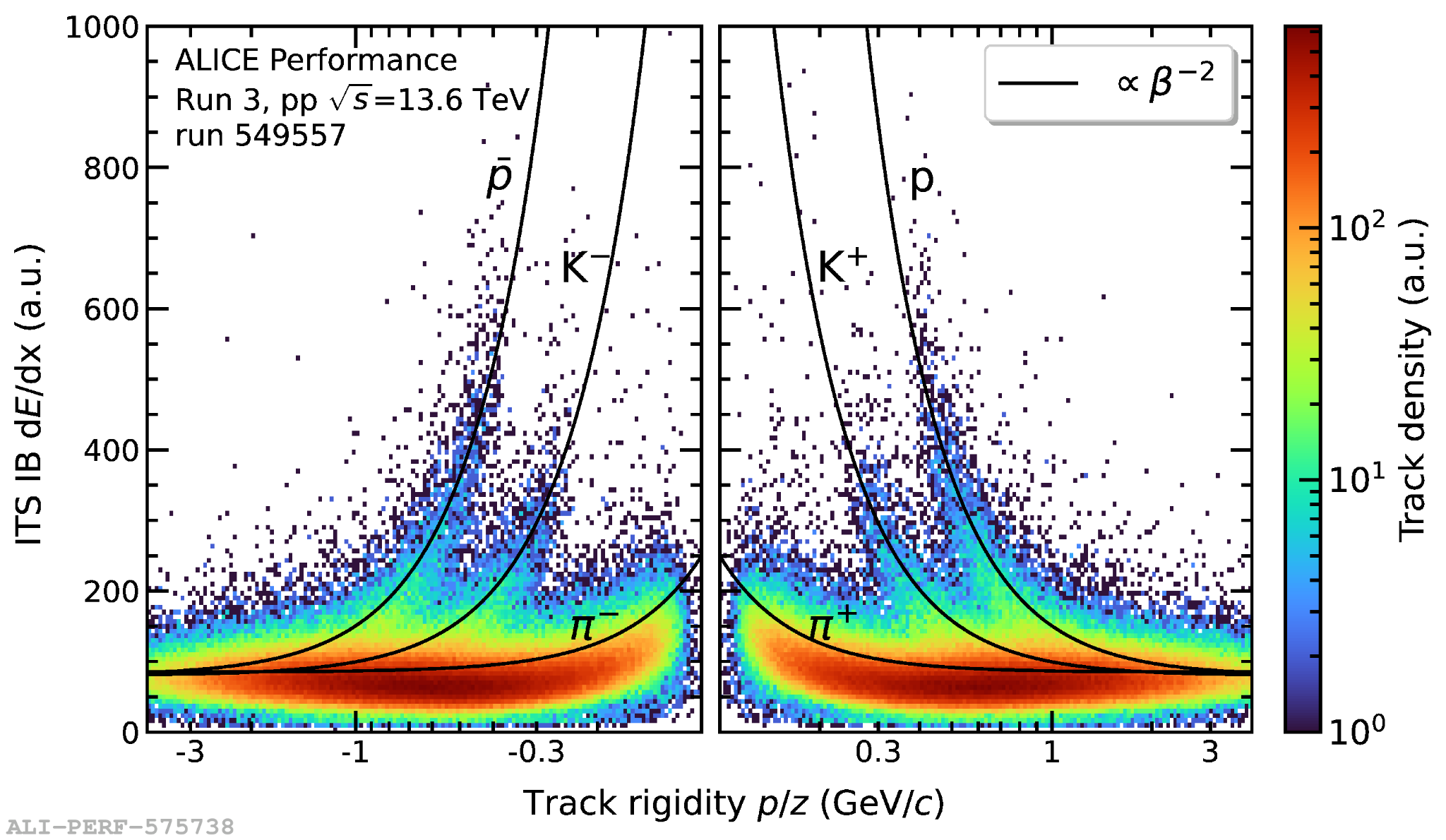}
\caption{d$E$/d$x$ spectrum versus track rigidity in the ALICE ITS2 Inner Barrel for $\sqrt{s} = 13.6$ TeV pp collisions with interaction rate $\sim$1 kHz. Taken from \cite{A_triolo_calib}. 
\label{fig:pid}}
\end{figure}

Recent studies of the time over threshold information from ALPIDE chips have revealed the potential for particle identification using MAPS. During a special data-taking, the ITS was configured to minimize signal clipping in the analogue front-end, allowing the signal length to be proportional to the deposited charge. Additionally, the signal was oversampled using an increased readout frame rate of 2.2 MHz ($\approx$ 11 times the nominal rate) to achieve high-resolution measurement of the Time-over-Threshold signal. To ensure stable detector performance under these conditions, data was recorded at a very low interaction rate, below 1 kHz of pp collisions. Offline reconstruction using a dedicated workflow, provided access to the d$E$/d$x$ spectrum as a function of track rigidity, based entirely on data from ALPIDE MAPS (see Figure \ref{fig:pid}). As orientation, reference lines proportional to $a \times \beta^{-2} + b$ were included. 

These studies represent the first proof of concept for particle identification using binary readout MAPS with a 25 $\mu$m-thick epitaxial layer.

\section{Summary}

ITS2 is the first fully MAPS-based tracking detector at the LHC and represents the largest and most granular pixel sensor successfully operated in both proton-proton  and lead-lead collisions. During Run 3, the ITS2 recorded over 80 pb$^{-1}$ of pp collision data at an interaction rate of 500 kHz and 2.16 nb$^{-1}$ of Pb--Pb data at 50 kHz. Throughout this period, the detector demonstrated stable performance, with less than 1\% loss of acceptance due to errors. The experience gained from Run~3 operations provides valuable insights for the development of ITS3 and ALICE3.



\begin{thebibliography}{99}

\bibitem[]{LS2}
S.~Acharya \textit{et al.} [ALICE],
JINST \textbf{19} (2024), P05062,
doi:10.1088/1748-0221/19/05/P05062.


\bibitem[]{its2}
B.~Abelev \textit{et al.} [ALICE],
J. Phys. G \textbf{41} (2014), 087002,
doi:10.1088/0954-3899/41/8/087002.

\bibitem[]{its2}
B.~Abelev \textit{et al.} [ALICE],
J. Phys. G \textbf{41} (2014), 087002,
doi:10.1088/0954-3899/41/8/087002.

\bibitem[]{alpide}

G.~Aglieri Rinella [ALICE],
Nucl. Instrum. Meth. A \textbf{845} (2017), 583-587,
doi:10.1016/j.nima.2016.05.016.

\bibitem{A_triolo_calib}
A.~S.~Triolo [ALICE],
JINST \textbf{20} (2025) no.01, C01030
doi:10.1088/1748-0221/20/01/C01030
[arXiv:2409.19810 [physics.ins-det]].


\bibitem[]{radiation_efficiency}
S.~Kushpil, F.~Krizek and A.~Isakov,
IEEE Trans. Nucl. Sci. \textbf{66} (2019), 2319-2323,
doi:10.1109/TNS.2019.2945234.


\bibitem[]{hillemanns2013}
H.~Hillemanns \textit{et al.},
2013 IEEE Nuclear Science Symposium and Medical Imaging Conference (2013 NSS/MIC), Seoul, Korea (South), 2013, pp. 1–5,  
doi:10.1109/NSSMIC.2013.6829475.





\end{thebibliography}


\end{document}